\documentclass{article}
\usepackage{spconf,amsmath,graphicx}

\usepackage{cite}  
\usepackage{amssymb,amsfonts}
\usepackage{algorithmic}
\usepackage{graphicx}
\usepackage{textcomp}
\usepackage{xcolor}
\usepackage{multirow}
\usepackage{hyperref}
\usepackage{enumitem}

\newcommand{\R}{\boldsymbol{R}}
\newcommand{\T}{\boldsymbol{T}}
\newcommand{\X}{\boldsymbol{X}}
\newcommand{\bTheta}{\boldsymbol{\Theta}}

\title{Modurec: Recommender Systems with Feature and Time Modulation}
\name{Javier Maroto, Clément Vignac and Pascal Frossard}
\address{EPFL, Switzerland}

\begin{document}
\ninept
\maketitle

\begin{abstract}

Current state of the art algorithms for recommender systems are mainly based on collaborative filtering, which exploits user ratings to discover latent factors in the data. 
These algorithms unfortunately do not make effective use of other features, which can help solve two well identified problems of collaborative filtering: cold start (not enough data is available for new users or products) and concept shift (the distribution of ratings changes over time).
To address these problems, we propose Modurec: an autoencoder-based method that combines all available information using the feature-wise modulation mechanism, which has demonstrated its effectiveness in several fields.
While time information helps mitigate the effects of concept shift, the combination of user and item features improve prediction performance when little data is available.
We show on Movielens datasets that these modifications produce state-of-the-art results in most evaluated settings compared with standard autoencoder-based methods and other collaborative filtering approaches.

\end{abstract}


\section{Introduction}
Recommender systems seek to suggest items that users are susceptible to like, given some information about users, items and the history of previous ratings. They are of high interest due to their primary use in commercial applications like marketing or product recommendation.

Recommender systems are based on one of the following two strategies. \emph{Content filtering} \cite{belkin1992information,lang1995newsweeder,pazzani1996syskill}
leverages user and item features while \emph{collaborative filtering} \cite{resnick1994grouplens,billsus1998learning,yu2004probabilistic}
only relies on the past ratings.
In practice, collaborative filtering tends to perform better than content filtering because of the richness of the rating history, which allows to discover latent features in the data and to go beyond what most user and item features can offer \cite{glauber2019collaborative}.
However, this dependency on the rating history often leads to the \emph{cold start problem} \cite{mansur2017review}: recommendations to new users (or of new items) may not be good due to the lack of sufficient ratings, thus holding back new users from joining the platform. On the contrary, for users that have rated items for a long time, predictions might not be accurate anymore because they are based on old ratings, which leads to the problem called \emph{concept drift} \cite{schlimmer1986beyond, widmer_learning_1996}. 
To tackle these issues, the authors in \cite{rendle_difficulty_2019} proposed to use the rating timestamps, which are easily available.
They transform the rating date into a categorical variable and concatenate it with the rest of the data. These feature vectors are passed to an algorithm that generalizes matrix factorization to multiple inputs \cite{rendle2012factorization}.

In this paper, we propose a new method to solve both cold start and concept drift within the autoencoder framework, which has been shown to give very good results despite its simplicity \cite{sedhain_autorec:_2015}. Differently than previous solutions, we scale and process the timestamps directly. We then use modulation instead of concatenation, as it avoids to explicitly learn all pairwise interactions between the time and the rest of the features. This approach results in better performance than competitor methods with fewer parameters.

Formally, for a recommendation system with $M$ users and $N$ items, we have to fill a $M \times N$ matrix $\hat \R$ given the rating history $\R \in \mathbb R^{M \times N}$, its timestamps $\T \in \mathbb R^{M \times N}$ and relevant user $\X_u \in \mathbb R^{M \times d_u}$ and item features $\X_i \in \mathbb R^{N \times d_i}$.
In order to use all available data, we propose to condition the ratings on side information using feature-wise transformations \cite{perez_film:_2018} which allow us to learn non-linear trends in a end-to-end fashion. This can be done with only very few parameters, thus reducing the risk of overfitting.
We also find that user and rating features are mostly useful when little data is available and detrimental otherwise. As a result, we propose to use them to modulate the ratings in an adaptive fashion, according to the number of ratings that a user has already given.
The modulated ratings are then used within an autoencoder framework, which has proved to be a simple and yet very effective way to perform collaborative filtering and solve recommendation problems \cite{sedhain_autorec:_2015}.

We show through experiments that our new method achieves state-of-the-art results on the MovieLens 100K and 1M regression tasks, which confirms that it is important to use all available information in the recommendation task. Ablation studies show that the performance gain mostly comes from time information. Also, the user and item features significantly help for users with few ratings, when added adaptively. We thus advocate to not only use the rating matrix but all available information, and to use modulation mechanisms to combine all sources together in a simple and computationally efficient way.

The rest of the paper is organised as follows. We describe related work in Section \ref{sec:related_work}. In Sections \ref{sec:coll_filt_ft_mod} and \ref{sec:implementation}, we explain our proposal and detail our model design choices. Finally, in Section \ref{sec:experiments}, we test our approach with ablation studies and discuss our experimental results.

\begin{figure*}[ht]
    \centering
    \includegraphics[width = 0.9\linewidth]{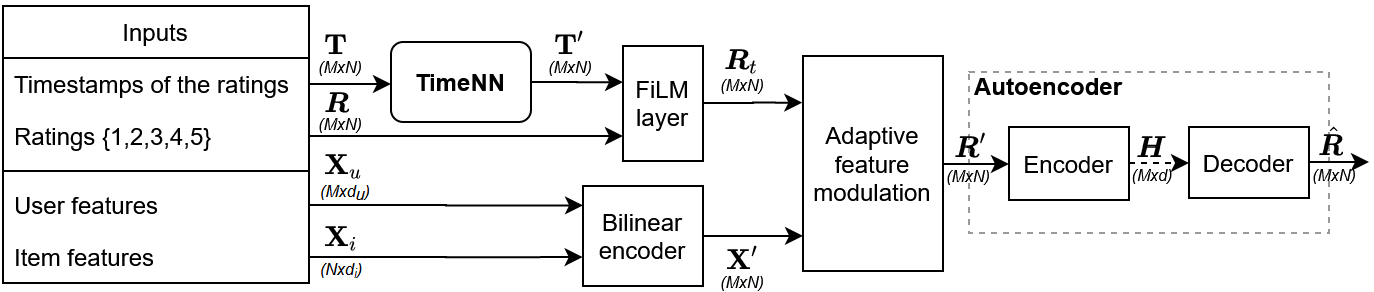}
    \caption{Modurec architecture. The time is processed with the TimeNN module and combined with the ratings using a FiLM layer. The feature information is added based on the sparsity of the time and rating data. Finally, the autoencoder process all data and fills the rating matrix.}
    \label{fig:model}
\end{figure*}

\section{Related work}
\label{sec:related_work}
Collaborative filtering methods for recommender systems can be essentially divided into matrix factorization, autoencoder-based methods and geometric matrix completion.
\emph{Factorization machines} \cite{koren2009matrix, xiong2010temporal,rendle2010pairwise}
represents one of the earliest and most popular approaches in recommender systems. 
They are a generalization of matrix factorization problems \cite{srebro2003weighted} for multiple features and n-wise low-rank interactions.
The authors in \cite{rendle_difficulty_2019} showed state-of-the-art results with a factorization machine that learns pairwise interactions between ratings, user, item and time features. One of the main drawbacks of this approach is the high number of parameters required to model all these interactions.

The assumption that the data can be represented in a compact way in a latent space is also the motivation for autoencoder-based models \cite{ballard1987modular, sedhain_autorec:_2015}. Autoencoders are composed of an encoder that projects the rating matrix $\R$ of size $M$x$N$ into a smaller embedding matrix of size $M$x$d$, and a decoder, whose objective is to reconstruct the original matrix.
Because of the inductive bias of the network and the information bottleneck in the latent space, the partially observed matrix is not reconstructed exactly and zero entries are replaced by a prediction for unobserved ratings. With respect to matrix factorization methods, autoencoders have the advantage of being inductive, which means that one can make predictions for new users without having to retrain the model \cite{zhang2019inductive} -- this is a key requirement for real systems.
Autoencoders of various complexities have been proposed: while the work in \cite{sedhain_autorec:_2015} simply uses a linear layer for both the encoder and the decoder, the authors in \cite{strub2016hybrid} propose to connect additional user/item features to the autoencoder and the study in \cite{muller_kernelized_2018} introduces the concept of kernelized weight matrices, which reduces the number of free parameters at the cost of additional hyperparameters like the choice of the kernel function.

Finally, with the recent popularity of geometric deep learning methods \cite{bronstein2017geometric}, several works have studied the reconstruction of the ratings using graph neural networks \cite{berg_graph_2017, monti_geometric_2017, zhang_star-gcn:_2019}. 
The underlying assumption of these methods is that the input has a set structure with no natural ordering, which is the case of users and items in a recommender system. 
However, results show that these methods fall behind simpler models \cite{rendle_difficulty_2019}, which may be explained by the difficulty to build good graphs to represent the data: since the node degree distribution typically follows a power law, very popular nodes are over-represented during training.

The above collaborative filtering methods typically only leverage the historic ratings. We rather propose in this work to use a feature-wise modulation mechanism \cite{perez_film:_2018} to improve performance in cold-start scenarios. Such mechanisms have been applied with great success to many other tasks where various sources of information need to be combined: in visual reasoning \cite{perez_film:_2018}, which combines linguistic and visual information, image synthesis \cite{brock2018large}, where modulation is used to project class embeddings onto convolutional layers, and few-shot learning \cite{oreshkin2018tadam}, where it is used to condition a feature extractor with task embeddings.

\section{Collaborative filtering with feature modulation}
\label{sec:coll_filt_ft_mod}
We now present Modurec, a new end-to-end model which is able to use various input sources within an inductive collaborative filtering framework. 
In this work, we assume access to an (incomplete) rating matrix $\R \in \mathbb R^{M \times N}$, timestamps $\T \in \mathbb R^{M \times N}$ for these ratings as well as some user features $\X_u \in \mathbb R^{M \times d_u}$ and item features $\X_i \in \mathbb R^{N \times d_i}$.
The core idea of Modurec is to first compute an intermediate matrix
$\R' = f(\R, \T, \X_i, \X_u) \in \mathbb R^{M \times N}$ that stores all available information, and then to use it as input to a standard autoencoder that will produce a prediction matrix $\hat \R = g(\R') \in \R^{M \times N}$. That way, the autoencoder can learn latent features that depend on all information and not only the rating data. The Modurec architecture is summarized in Figure \ref{fig:model}. 

In order to parametrize the function $f$, we rely on the feature-wise modulation mechanism: first, we apply the same multi-layer perceptron called TimeNN, to each timestamp $\T_{ij}$ to capture non-linear trends in the data in such a way that the number of parameters does not grow with $M$ or $N$. The resulting matrix $\T'$ is then used to modulate the ratings using a FiLM layer \cite{perez_film:_2018}, which computes
$\R_t = \alpha \R + \beta \T' + \gamma \R \cdot \T'$,
where $\alpha$, $\beta$, $\gamma$ are trainable scalars.
This mechanism allows to combine both additive and multiplicative modulations at the cost of very few additional trainable parameters. 

In order to use the item and user features to modulate $\R_t$ in a similar fashion, we need to convert $\X_u$ and $\X_i$ into a $M \times N$ feature matrix $\X'$. To do so in a simple way, we propose to use a bilinear encoder: $\X' = \X_{i}\boldsymbol{\Theta} \X_{u}^{T}$ where $\bTheta \in \mathbb R^{d_u \times d_i}$ is a matrix of trainable weights. Yet, one can observe that predicting ratings from user and items features is precisely the content filtering problem, so that any content filtering method could actually be used here.

One key element of our design relies in the proper combination of the processed features $\X'$ with the matrix $\R_t$ containing the ratings modulated by time information. Experimentally, we observed that trivially combining the processed features and the ratings does not work. This can be explained by the fact that features are not equally relevant to all users: the fewer ratings we have, the more the system should rely on the processed features. Thus, we propose a novel adaptive approach that weights both information sources entry-wise based on the number of ratings that each user/item has given/received. This method is quite effective experimentally.

Finally, once all different data sources are combined, the resulting $\R'$ matrix is passed through the autoencoder, of latent dimensionality $d$, which will use the contextual data to reconstruct the original $\R$. The model is trained by minimizing L2 loss between the predictions and the true ratings $\R$. For the autoencoder  we simply use Autorec \cite{sedhain_autorec:_2015}, because of its simplicity and good performance. The details of our implementation are provided in the next Section.

\section{Modurec implementation}
\label{sec:implementation}


\subsection{Processing and modulating the time information}
\label{ssec:time}

In recommendation platforms, rating events are recorded with timestamps,
which can be used to discover trends in the data. These trends can be global (e.g., people may become more critical over time and give on average lower ratings), but they may also be specific to a user or a movie. For example, movies that raised criticism at first may become cult over time. To be able to model these different aspects, we first map the timestamps to three values: 
\begin{itemize} 
    \item \emph{Global}: time relative to the creation of the platform.
    \item \emph{User}: time relative to the user first rating.
    \item \emph{Movie}: time relative to the movie first rating.
\end{itemize}
These time values are then normalized to $[0, 1]$, and used as input to the neural network TimeNN. We parametrize this network with three fully-connected layers with hidden size (3, 32) and ReLU nonlinearities. The output of the TimeNN is one value for each timestamp, resulting in a $M$x$N$ matrix, namely $\T$.

\subsection{Processing and modulating the user and time features}
\label{ssec:fts}


In order to obtain the intermediate matrix $\R'$ with all the available information, we need to combine the processed features $\X'$ and the time-modulated rating data $\R_t$. We use a weighted matrix $\textbf{A}$ that regulates the relative importance of each of the two matrices for each entry:
\begin{equation}
\label{eq:adaptive_combiner}
    \R' = \textbf{A} \cdot \R_t + (1 - \textbf{A}) \cdot \X'
\end{equation}
where each matrix element $\textbf{A}_{ij}$ is given by
\begin{equation*}
\textbf{A}_{ij} = 
   \begin{cases}
   \sigma (w_1 |\mathcal{O}_{i,i}| + w_2 |\mathcal{O}_{u,j}| + b) &|\mathcal{O}_{i,i}|, |\mathcal{O}_{u,j}| > 0\\
  0 &|\mathcal{O}_{i,i}|, |\mathcal{O}_{u,j}| = 0\\
   \end{cases}
   \label{eq:A_matrix}
\end{equation*}
and $|\mathcal{O}_{i,i}|$ is the number of ratings of item i, $|\mathcal{O}_{u,j}|$ is the number of ratings of user j, $\sigma$ is the sigmoid function and $w_1$, $w_2$, $b$ are scalar trainable parameters. In general, $\textbf{A}_{ij} \approx 1$ since the data is mostly comprised of ratings of popular items, but for cold start we will only use the feature data, i.e., $\textbf{A}_{ij} = 0$.

\subsection{Autoencoder design}
\label{ssec:autoencoder}


For the autoencoder, we use the Autorec model \cite{sedhain_autorec:_2015}, which has a linear layer for both the encoder and the decoder. For the encoder, the activation function is a sigmoid, while for the decoder no activation function is used. Both functions are defined as follows: 
\begin{align*}
\begin{aligned}
&\boldsymbol{H} = \sigma(\R'\mathbf{W}_{enc} + \boldsymbol{b}_{enc})\\
&\boldsymbol{\hat{R}} = \boldsymbol{H}\mathbf{W}_{dec} + \boldsymbol{b}_{dec}
\end{aligned}\label{eq:autoencoder}
\end{align*}
where $\boldsymbol{H}$ is the latent factor matrix of shape $M$x$d$, $\boldsymbol{\hat{R}}$ is the predicted rating matrix of shape $M$x$N$, and $\mathbf{W}_{enc}$, $\mathbf{W}_{dec}$, $\boldsymbol{b}_{enc}$ and $\boldsymbol{b}_{dec}$ are trainable weight matrices and bias vectors of shape $N$x$d$, $d$x$N$, 1x$d$ and 1x$N$, respectively.


We made some changes to the Autorec model to address overfitting.
We found that adding dropout \cite{srivastava2014dropout} in the input and embedding layers helps reduce overfitting \cite{hinton2012improving}.
Particularly, in the input layer, dropout has been shown to be advantageous in the context of denoising autoencoders \cite{vincent2008extracting}, which is our case as we "denoise" the sparse signal received by the user/item. Additionally, input dropout allows explicit minimization of the prediction error on unobserved ratings, as well as the reconstruction error on observed ratings.

As in Autorec, we can differentiate between I-Modurec and U-Modurec, where in the latter $\R'$ is transposed before feeding it to the autoencoder. Although in this paper we only take into consideration I-Modurec for clarity and performance sake, we believe U-Modurec is worth considering for tasks like clustering, where the autoencoder latent vectors are needed for both items and users. Furthermore, ensemble methods can be used based on these two models, but we found experimentally that they do not improve performance.

\begin{table}[t!]
    \centering
    \caption{Complexity analysis of the forward pass during training for Modurec and its competitors. $M$ and $N$ denote the number of users and items, $T_r$ the number of training ratings, $d$ the embedding dimension, $V_r$ the maximal number of ratings for an item, $d_u$ and $d_i$ the number of user and item features. All notations are big-O.}\bigskip
    
    \begin{tabular}{c|cc}
        Method & Time & Memory \\
        \hline
        TimeSVD++ \cite{rendle2012factorization}& $T_r~d~(d_u + d_i)$ & $T_r$\\
        GC-MC \cite{berg_graph_2017}& $T_r d^2 + (N d_i + Md_u)d $ & $(N + M)d$\\ 
        \hline
        Modurec (dense)& $NM(d + d_i) + Md_u d_i$ & $M ~ d$\\ 
        Modurec (sparse)& $T_r(d + d_i) + Md_ud_i$ & $V_r ~ d$\\ 
    \end{tabular}
    \label{tab:complexity_comp}
\end{table}

\subsection{Complexity analysis}
\label{ssec:comp_anal}

The computational complexity of Modurec and its competitors is presented in Table \ref{tab:complexity_comp}.
Our method benefits from the advantages of autoencoder architectures, which are an easy GPU parallelization and a time complexity that is linear in both the training set size $T_r$ and the latent dimension $d$. Whereas we found an implementation with dense tensors to be sufficient for our experiments, a more efficient implementation would require two extra steps: first, write the autoencoder with sparse tensors and load the weight matrices only partially. Second, since $\X'$ is dense, we need to sparsify $\R'$. Experimentally, we observe that $\textbf{A}$ is close to 1 for most entries, so that we can threshold its values and obtain a sparsity structure for $\R'$ close to that of $\R$. Overall, we find that our method has significantly lower memory complexity than its competitors (as it does not depend on $N$) and that a sparse implementation would compare very favorably in terms of time complexity.




\begin{table*}[ht]
    
    \centering
    \caption{Average RMSE recommendation results on several datasets. We use Modurec\_[DFT] as the nomenclature for our model.\newline D = with dropout; F = with user and item features module; T = with time module.\newline The $^{*}$ sign refers to our own re-implementation.}\bigskip
    
    \addtolength{\tabcolsep}{-2pt}
    \begin{tabular}{c|cccccccc|ccc}
        Dataset & GRALS & sRGCNN & GC-MC & STAR & CF-NADE & Sparse & TimeSVD++ & I-Autorec$^{*}$ & Modurec & Modurec & Modurec\\
        & & & & -GCN &  & FC & flipped$^{*}$ & & \_D & \_DT & \_DFT\\
        \hline
        ML-100K & 0.945 & 0.929 & 0.905 & 0.895 & --- & --- & 0.890 & 0.908 & 0.905 & \textbf{0.887} & \textbf{0.887}\\
        ML-1M & --- & --- & 0.832 & 0.832 & 0.829 & 0.824 & 0.842 & 0.831 & 0.826 & \textbf{0.821} & \textbf{0.821}\\
        ML-10M & --- & --- & 0.777 & 0.770 & 0.771 & 0.769 & \textbf{0.749} & 0.782 & 0.789 & 0.777 & 0.779\\
    \end{tabular}
    \label{tab:results}
\end{table*}

\begin{table}[ht]
    \centering
    \caption{Effect of adding time information on the RMSE for SVD++ flipped and Modurec\_D (ours). The $^{*}$ refers to our implementation.}\bigskip
    
    \begin{tabular}{c|ccc}
        Dataset & Model & Without time & With time \\
        \hline
        \multirow{2}*{ML-100K} & SVD++ flipped$^{*}$ \cite{rendle_difficulty_2019} & 0.894 & 0.890\\
        & Modurec\_D & 0.905 & \textbf{0.888}\\
        \hline
        \multirow{2}*{ML-1M} & SVD++ flipped$^{*}$ \cite{rendle_difficulty_2019} & 0.843 & 0.842\\
        & Modurec\_D & 0.826 & \textbf{0.821}\\
    \end{tabular}
    \label{tab:timeinfo_comp}
\end{table}

\begin{table}[ht]
    \centering
    \caption{Effect of combining the user and item features of the \emph{Static} and the \emph{Adaptive} algorithms versus the \emph{Nothing} algorithm, for users and items that have "few ratings" or "many ratings", as defined in Subsection \ref{ssec:adaptive}.}\bigskip
    
    \begin{tabular}{c|ccc}
        Dataset & Algorithm & Few ratings & Many ratings \\
        \hline
        \multirow{3}*{ML-100K} & Nothing & 1.6093 & \textbf{0.8371}\\
        & Static & 1.6000 & 0.8417\\ & Adaptive & \textbf{1.3412} & 0.8380\\
        \hline
        \multirow{3}*{ML-1M} & Nothing & 1.1481 & \textbf{0.7895}\\
        & Static & 1.1457 & 0.7900\\ & Adaptive & \textbf{1.1360} & 0.7897\\
    \end{tabular}
    \label{tab:feature_comp}
\end{table}

\section{Experiments}
\label{sec:experiments}

\subsection{Settings}

We now analyze the performance of Modurec\footnote{Code: \href{https://github.com/LTS4/modurec}{https://github.com/LTS4/modurec}} and compare it with the relevant literature. 
We use the MovieLens 100K, 1M and 10M datasets \cite{harper2016movielens}. These are well known and widely used datasets for benchmarking recommender algorithms. Whereas the dataset is built in such a way that all users have at least 20 ratings, they have some items with no ratings so that strict cold start is present in the datasets. One additional property that differentiates them from the other datasets is the inclusion of timestamps in the ratings, which allows us to show the importance of using timestamps as extra features.


For ML-100K, we use the first of the five provided data splits (80\% training, 20\% testing) like most methods of the literature. For ML-1M and ML-10M, we apply out-of-sample splitting for the observed ratings and timestamps using 90\% for training and 10\% for testing. In both datasets, we use 10\% of the training set as hold-out to tune all hyperparameters. As for the metric, we compute the RMSE between the test ratings (valued from 1 to 5) and our predictions. Most works in the relevant literature use the same settings, namely state-of-the-art matrix factorization algorithms like GRALS \cite{rao2015collaborative} and TimeSVD++ flipped \cite{rendle_difficulty_2019}; geometric matrix completion algorithms like sRGCNN \cite{monti_geometric_2017}, GC-MC \cite{berg_graph_2017}, STAR-GCN \cite{zhang_star-gcn:_2019} and CF-NADE \cite{zheng2016neural}; and autoencoder-based algorithms like Autorec \cite{sedhain_autorec:_2015} and Sparse FC \cite{muller_kernelized_2018}. We have chosen $d = 500$ for the Modurec's autoencoder as in the original Autorec implementation.

\subsection{Performance Analysis}


Table \ref{tab:results} summarizes the results of Modurec (in the table as Modurec\_DFT) and its ablated versions compared to its competitors. If not stated otherwise, the results are directly taken from the respective papers for the competitor methods. We report the average results over 10 runs. For ML-1M, the splits are randomized between experiments. Because of time constraints, we compute the ML-10M experiment once. Although Autorec is already a very strong baseline when input dropout is added, our full model outperforms it significantly. While Modurec yields state of the art results for ML-100K and ML-1M, the higher sparsity of the ML-10M rating is detrimental to basic autoencoder models, which explains the reduced performance.


The results in Table \ref{tab:results} highlight the importance of addressing concept shift in recommender systems. This can be done using modulation, or by simply concatenating the time features to the ratings as in \cite{rendle_difficulty_2019}. In Table \ref{tab:timeinfo_comp}, we compare these two approaches: we find that modulation performs much better, thanks to its ability to learn non linear trends with very few parameters.




\subsection{Adaptive user and item feature modulation}
\label{ssec:adaptive}

The purpose of doing feature modulation is to overcome the shortcomings of collaborative filtering, namely cold start, without compromising its performance in other settings. From the results in Table \ref{tab:results}, it seems that our feature modulation module has no impact. However, this is explained by the fact that the datasets have been built in such a way that user cold start is avoided and the testing uses out-of-sample sampling, where most ratings come from users and items with a high number of ratings. In practical applications, out-of-time sampling makes more sense and cold start has a much higher impact in the results. 

To properly measure performance under cold start, we selected two subsets of ratings. The "few ratings" group represent the subset of ratings where both $|\mathcal{O}_{i,i}|$ and $|\mathcal{O}_{u,j}|$ are in the bottom quantile, while the "many ratings" group is the subset of the top quantile. The latter is used to control that improving performance on cold start does not jeopardize the other cases. For reference, they represent a 0.8 \% and a 28.7 \% of the ML-100K ratings, respectively. We test the following three versions of Modurec:
\begin{itemize}
    \item \textbf{Nothing}: Modurec without the bilinear encoder and the adaptive combiner (no user or item features are used).
    \item \textbf{Static}: Modurec, where we have a much simpler combiner instead of our adaptive combiner. It is characterized by the following relation: $\R' = \alpha \R_t + (1 - \alpha) \X'$, where $\alpha$ is a scalar trainable parameter.
    \item \textbf{Adaptive}: Modurec, with its adaptive combiner as in Eq \eqref{eq:adaptive_combiner}.
\end{itemize}
We compare these three versions in Table \ref{tab:feature_comp}, where we see that for both datasets we have similar behaviours. The \emph{Nothing} algorithm performs best when we have many ratings, as expected from pure collaborative filtering models. The \emph{Static} algorithm outlines the limitations of combining user and item features in a fixed ratio. The lower the $\alpha$ value, the better it performs when there are few ratings, but at the cost of lower performance when we have more ratings. As a result, the model converges to a very high $\alpha$ (in our testing, around 0.97), which not only does not solve cold start, but reduces general performance. Our \emph{Adaptive} algorithm does not have that limitation, since $\textbf{A}_{ij}$ can even be zero on cold start but close to one for all other cases. This is even more evident on ML-100K, where we have less total ratings per user/item.

\section{Conclusion}

In this work, we have shown the importance of properly using the time information present in recommender systems. We also show how to effectively tackle the cold start problem with a novel adaptive algorithm that is agnostic of the content and the collaborative filtering models in use. 
We believe that data-driven approaches should be given more importance, and we recommend trying adaptive weighting methods when combining conflicting sources of data and FiLM modulations when the sources complement each other.

\bibliographystyle{IEEEbib}
\bibliography{references}
\end{document}